# The non-linear behavior of aqueous model ice in downward flexure

## Authors:

*Rüdiger U. Franz von Bock und Polach (Hamburg University of Technology)\**
*Robert Ettema (Colorado State University)*
*Silke Gralher (Hamburg University of Technology)*
*Leon Kellner (Hamburg University of Technology)*
*Merten Stender (Hamburg University of Technology)*

\*Corresponding author: franz.vonbock@tuhh.de

**Abstract**

As aqueous model ice is used extensively in ice tanks tests on the performance of ship hulls in sheet ice, it is imperative that such model ice replicate the main flexural strength behavior of sheets of sea ice and freshwater ice. Ice tanks use various types of aqueous model ice types, each of which contain brine dopants to scale-reduce ice-sheet strength. Dopants, though, introduce non-linear trends in the scaled flexural behavior of model ice sheets, and can affect ice loads and ice-rubble at ship-hulls and structures. This paper analyzes the non-linear behavior of model ices, and shows that all types behave non-linearly in flexure independent from crystal structure or chemical dopant. Such behavior is attributable to plasticity and vertical variations in stiffness and strength through sheets of model ice. Additionally, the problematic formation of a top layer in model ice sheets is shown to have a greater impact of sheet behavior than the literature reports heretofore. There remains a significant knowledge gap regarding the freezing and movement of brine dopants within ice sheets and their impact on the non-linear behavior. Additionally, it is found that the Hertz method for estimating the Cauchy number of model ice does not reflect the actual deformation behavior of model ice and should be revised.

## 1. Introduction

Physical modelling is commonly used to assess the performance of ship hulls moving through level sheets of ice. Such modeling involves concepts of similitude applied, using an ice tank, to a moving ship hull that loads an ice sheet predominantly in downward flexure, displacing the resulting pieces of broken ice as the hull proceeds through the sheet ice (Vance 1975, Schwarz 1977, Tatinclaux 1988). *Figure 1* shows a sheet of aqueous model ice failing beneath the bow of an icebreaker hull. The general convention adopted for such physical modelling (e.g., Schwarz 1977, ASCE 2000) is that suitably controlled model ice can mechanically reproduce the flexural behavior of first-year sheets of ice. However, aqueous model ice (water-based model ice) is subject to a non-linear strength behavior that can adversely affect the model ice's capacity to replicate the strength behavior of prototype level ice. The non-linear behavior of model ice entails ice-sheet deformations that significantly exceed those associated with prototype-ice behavior. Modelers, aware of this concern, have suggested limits in modeling scale and procedure, and generally tried to improve the flexural behavior of model ice sheets (e.g., Schwarz 1977, Enkvist and Mäkinen 1984, Timco 1986, von Bock und Polach 2015).

Excessive flexural deformation is a scale effect that increases model-scale deformation prior to ice-sheet failure and, thereby leads to greater ride up of a ship hull onto an ice sheet. This effect increases the vertical motions of model hulls (Schwarz 1977, von Bock und Polach 2016). Vertical ship motions and the additional work needed to bend and displace the model ice sheet introduce an additional horizontal resistance component that increases model-scale resistance (Ettema et al. 1987, von Bock und Polach and Ehlers 2011). Model tests typically entail measurement and scaling of resistance forces, and observation of broken-ice along the hull (e.g., Li et al. 2017). If ice were too compliant and submerged in flexure under the hull (instead of breaking and creating ice rubble in front of the hull), the non-linear behavior of model ice inadequately reproduces the prototype behavior of a hull moving through an ice sheet. Essentially the same concern arises with current interests in physical modeling ice-sheet interaction with waves.

This paper reviews and evaluates the non-linear flexural behavior of aqueous forms of model ice, and explains how the structure of all such forms of model ice leads to non-linear flexural strength behavior. For this purpose, this paper analyzes data from ice sheets formed in the world's leading ice tanks: Aalto University, Aker Arctic, The Hamburg Ship Model Basin (HSVA), Krylov State Research Centre (KSRC) and Japan's National Marine Research Institute (NMRI). Besides discussing the physical processes involved in the flexural behavior of ice sheets, the evaluation uses a standard method, the Pearson correlation coefficient test, giving a quantitative measure of the strength of relationship between two variables.

To meet the similitude and scaling criteria specified for modeling ice-breaking hulls, modeling at these ice tanks follows the usual similitude consideration of reducing the flexural strength, $\sigma_f$, of prototype ice sheets in accordance with the length scale of the model hull, $\lambda$, (e.g., Vance 1975, Schwarz 1980, Ettema and Zufelt 1996). Further, modeling should encompass the important combinations of prototype ice thickness and flexural strength. Moreover, flexural deformation of the model ice should conform to values of the Cauchy number (ratio of inertia force/elastic-deformation force; $Ch = \rho V^2/E$) and the stiffness ratio $\sigma/E$ for the prototype ice; where $Ch =$

Cauchy number, $\rho$ = ice density, $V$ = a representative velocity (of the ship hull), $E$ = the modulus of ice sheet elasticity in flexure, and $\sigma_f$ = flexural strength of the ice sheet. The non-linear behavior of model ice in downward flexure failure, however, affects modeling similitude. This paper investigates whether such non-linear behavior is an inherent, general feature of model ice sheets; and indeed, if so, which sheet properties affect the behavior. The investigation requires understanding the structure of aqueous model ice, as described in the following section.

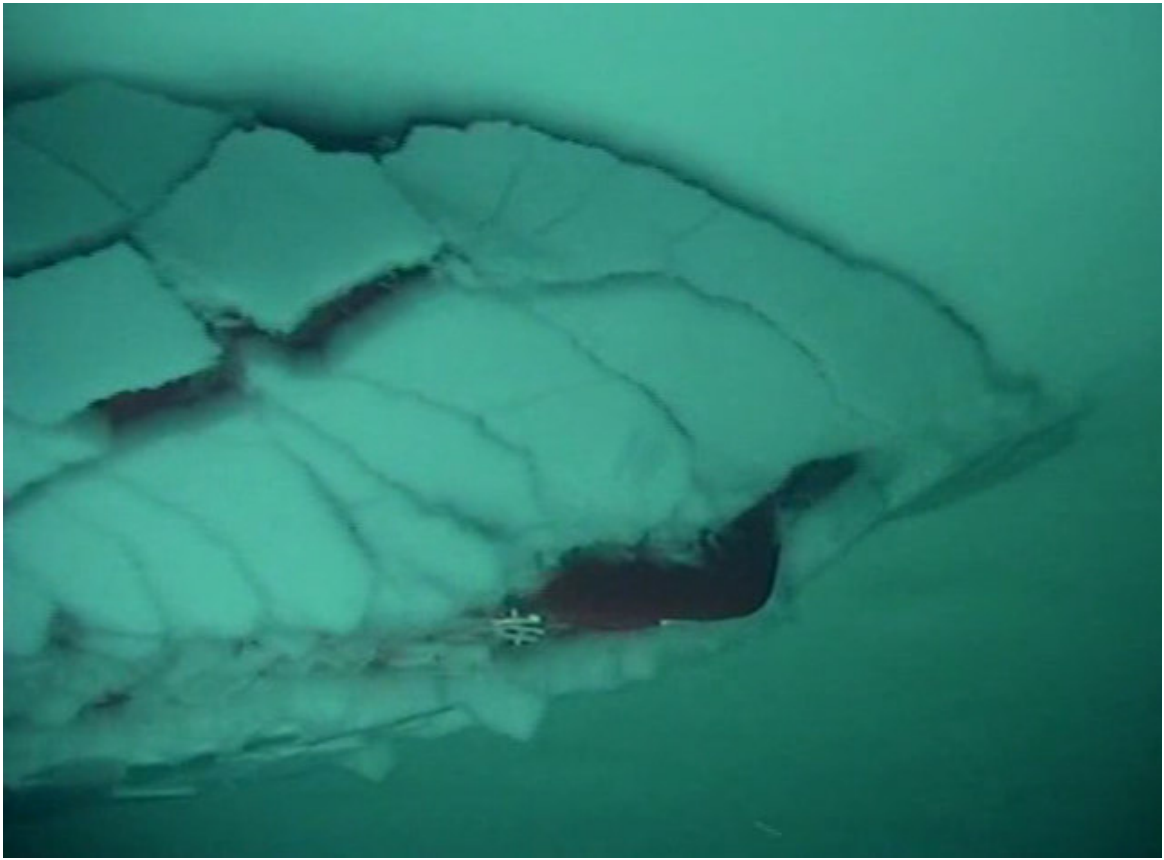

*Figure 1: Underwater view of ice-breaking pattern with model-ice failing in downward flexure (courtesy of HSVA)*

## 2. Aqueous Model Ice

The several forms of aqueous ice sheet that exist are classified in terms of the crystal structures used to adjust values of sheet flexural strength and stiffness. Additionally, the various dopants (chemical additives) used to create solutes contained within the sheets, thereby potentially affecting sheet strength, are mentioned in the classification.

Two types of ice-crystal structure are commonly distinguished. One type has a columnar structure, for which the model ice has an initial layer (top layer a few millimeters thick) of randomly oriented grains that then grow downwards in long thin columnar crystals (columnar ice, Figure 2a). The other one type is called fine-grained (FG) model ice (Figure 2b), where the model ice is accreted upwards by laminating layers of fine-grained ice over the water surface in an ice tank. FG model ice was initially developed by Enkvist and Mäkinen (1984) in an effort to address shortcomings in columnar model ice. Table 1 lists the ice tanks currently conducting physical modelling, and indicates the type of dopant and the crystal structure used as model ice. The high diversity in model ice used is attributable to each tank's history and the experience of its modelers (e.g., Enkvist and Mäkinen 1984, Timco 1986, Jalonen and Ilves 1990, Evers and Jochman 1993, Lee et al. 2010).

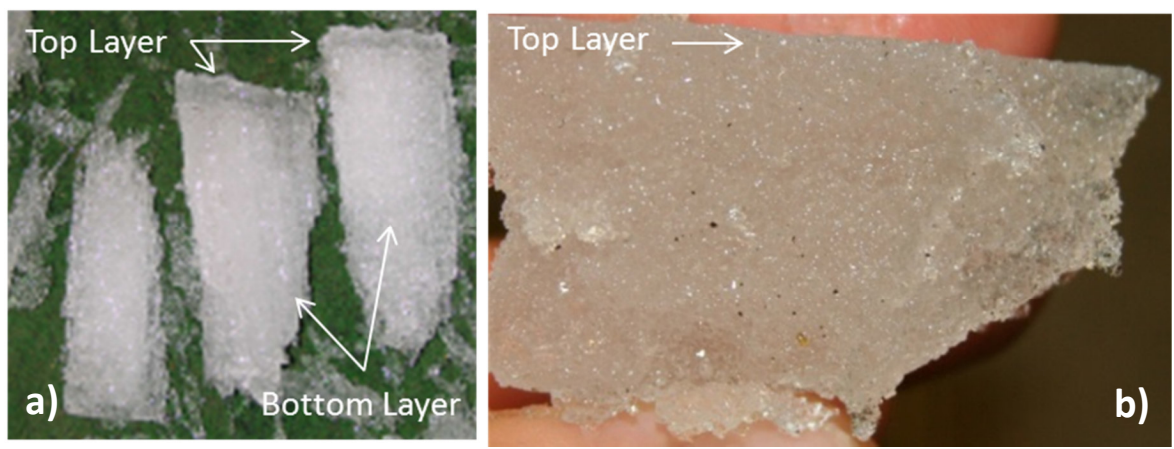

Both types of model ice involve water containing (or doped with) a weak solution of a chemical additive that acts as a solute into which ice crystals grow. The so-called doped model ices attain strengths substantially less than those observed for sea ice or freshwater ice, because the solute (dissolved dopant) entrapped within the crystal structure of the model ice weakens the overall ice structure. The crystal lattice of ice forming the structure does not readily incorporate solute molecules, but rejects them into the surrounding solution. As the diffusion of expelled dopant is relatively fast compared to the growth rate of ice-crystal, significant amounts of dopant are not entrapped in ice crystals (e.g., Timco 1981). Instead, pockets of enriched solute are trapped within the ice lattice. As the ice cools during growth, the pockets decrease in size and solute concentration increases within the solution in the pockets, thereby lowering its freezing temperature. However, upon tempering or warming to reduce ice-sheet strength, the solute pockets expand due to melting of the surrounding ice structure. The rate and degree of warming determine the strength properties of a test ice sheet, enabling modelers to reduce the sheet's strength to target values (e.g., Evers and Jochman 1993). The structure and, thereby, the strength behavior of model ice are affected by the type of dopant used, the extent of air inclusion, the temperature of the ice-basin water at the moment of seeding ice crystal-nuclei and by the air temperature during ice-sheet growth.

The first columnar model ice used salt (sodium chloride) as dopant, as this chemical occurs in sea ice (Shvayshteyn 1959). Today, most ice tanks use a dopant chemical of low molecular weight to trigger the internal melting (see Table 1). The ice tanks in Canada (NRC) and Korea (KRISO) use a multi-chemical dopant, with each chemical having a dedicated purpose. NRC uses the model ice developed by Timco (1986), where ethylene glycol (EG) is the low molecular chemical for the internal melting and aliphatic detergent (AD) is added to reduce the surface tension of the water layer at the ice interface to trigger the trapping of the low molecular weight chemical (EG) within the ice; i.e., to increase the efficiency of the amount of dopant used. The third component, not employed by KRISO is sugar, a chemical of high molecular weight that inhibits the lateral growth of crystals, thereby creating a fine-grained columnar ice.

For FG model ice, once the spray-crystals settle on the surface of the ice basin, water from the ice basin fills the voids between settled ice crystals, initially forming a composite of ice crystals and ice-basin water. In analogy to snow ice, the fine grained model ice traps a high amount of liquid at the crystal boundaries. During layer consolidation, additional freezing and generation of higher dopant concentration occurs in a manner similar to that for columnar ice, as mentioned above.

*Table 1:  Ice-modelling basins and types of aqueous model ice used (von Bock und Polach et al. 2013, Lee et al. 2010, Evers and Jochman 1993).*

| Ice Basin Facility | Country | Type of Aqueous Model Ice | |
| --- | --- | --- | --- |
| | | Crystal structure | Chemical dopant |
| Aalto University | Finland | Fine-grained (FG) | Ethanol |
| Aker Arctic | Finland | Fine-grained (FGX)[1] | Sodium Chloride |
| Japan Marine United (formerly, Universal Shipbuilding Corporation (NKK)) | Japan | Fine-grained | Urea |
| Krylov State Research Centre (KSRC) | Russia | Both fine-grained and columnar | Sodium Chloride |
| Maritime Ocean Engineering Research Institute (KRISO) | Korea | Columnar | Ethylene-Glycol-Aliphatic-Detergent (EG/AD) |
| National Research Council – Ocean, Coastal and River Engineering (NRC-OCRE) | Canada | Columnar | Ethylene-Glycol-Aliphatic-Detergent-Sugar (EG/AD/S) |
| National Maritime Research Institute (NMRI) | Japan | Columnar | Propylene Glycol |
| The Hamburg Ship Model Basin (HSVA) | Germany | Columnar | Sodium chloride |

---

[1] FGX is fine grained ice, where the bottom layers are sprayed with fresh water and the upper rest with tank water

| Tianjin University | China | Columnar | Urea |
| --- | --- | --- | --- |

## 3. Plasticity and Non-homogeneity of Model Ice in Flexure

### 3.1. Model ice plasticity

Several researchers (e.g., Schwarz 1977, von Bock und Polach 2015) identify plastic deformation in downward flexure as a concern to be addressed in the use of model ice sheets. Plastic deformation and remnant plasticity prevent sheets from completely breaking in the flexural brittle-elastic model, and thereby alter the temporal record of flexural failure, a major concern in ice modeling.

This concern had not gotten much attention until von Bock und Polach (2015) and von Bock und Polach and Ehlers (2013) showed that the mechanical behavior of FG model ice doped with ethanol can only be modeled as an elasto-plastic material model, with the yield strength at around 1%-5% of the flexural strength. This finding essentially refers to a bi-linear material behavior, whereby the yield strength marks the upper limit of the elastic modulus as determined by the deflection of an infinite plate (ITTC Guidelines 7.5-02-04-02, 2014); this approach is also called the Hertz-method (Hertz 1884). Figure 3 compares the simulation of an elastic cantilever beam and a cantilever beam with the elasto-plastic material mode. The elastic modulus in both simulations is derived from measurements (von Bock und Polach et al. 2013) and the elasto-plastic model shows good agreement with actual force-displacement measurements (von Bock und Polach 2015).

Sazonov and Klementeva (2011) investigated the flexural behavior of granular model ice at KRSC's old ice tank. They mention that the initial deformation (section "1" in Figure 4) refers to elastic deformation, whereas the subsequent deformation produces a non-elastic strain modulus.

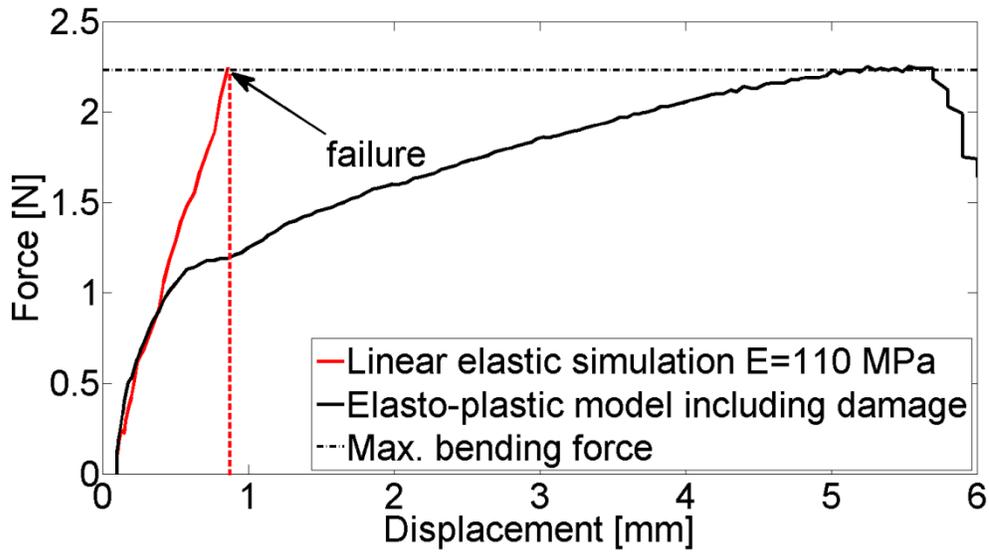

*Figure 3: Comparison of the simulation with the elasto-plastic model which represents the experiments and a linear elastic cantilever beam with properties derived from plate deflection tests (von Bock und Polach 2016).*

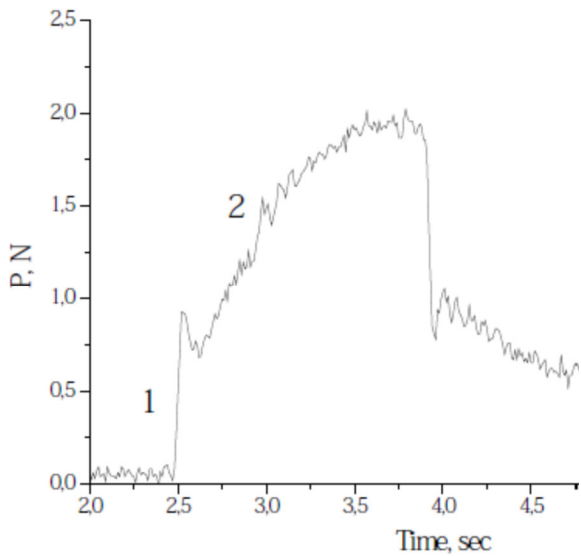

*Figure 4: Force-time progression of a cantilever beam test with granular ice at the Krylov State research Centre, adapted from Sazonov and Kelementeva (2011).*

Sazonov and Klementeva (2011) found that the sheet's flexural stiffness associated with *"1"* is 80%-85% (Table 2) of the elastic modulus determined using the Hertz method (ITTC Guidelines 7.5-02-04-02 2014). The stiffness in section *"2"* of Figure 4 is 50% (or less) of the initial stiffness, while the slope of the rest of the curve in Figure 4 is significantly smaller than that for section "2" and consequently also the stiffness and strain modulus.

Comparison of Figure 3 (simulation from von Bock und Polach 2015) and Figure 4 indicates a certain similarity in the flexural behavior of the two model ices (Aalto, FGX, 0.3% Ethanol; and, KSRI, FG, 1.5% NaCl). The beginning of the loading of the cantilever beam exhibits elastic deformation, and is followed by non-elastic deformation; i.e., both elastic and plastic deformations occur in the corresponding layers (von Bock und Polach, 2015).

*Table 2: Values of elastic modulus determined using the deflection of an infinite plate, and the based on the initial deformation section (section 1in Figure 4); from Sazonov and Klementeva (2011)*

| Ice thickness (mm) | Elastic modulus by deflection of an infinite plate / Hertz method (MPa) | Elastic modulus [MPa] based on stiffness in section "1" in *Figure 4* (MPa) |
|---|---|---|
| 33 | 32.1 | 27.3 |
| 43 | 63.9 | 51.4 |

Gralher (2017) analyzed the flexural behavior of HSVA's columnar model ice produced using sodium chloride (NaCl) as dopant. This model ice produced a non-linear force-displacement behavior, as Figure 5 shows for a piecewise linear representation of the measurement; here, $S_i$ is the slope of the section and $r$ is the ratio between the section slope and slope of the linear approach. The linear approach is a simplified method to define the effective stiffness or stain modulus in downward bending, which a ship model would experience in its interaction with the ice sheet. The initial slope of the force displacement curve of a cantilever beam test shows essentially elastic deformation of the sheet.

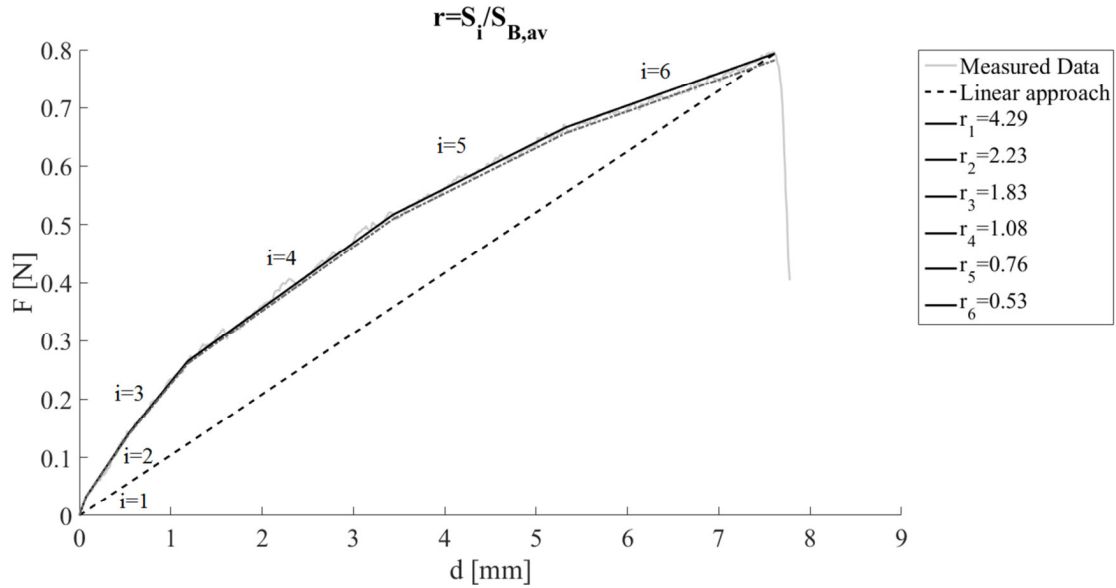

*Figure 5: Piece-wise linearization of the load deflection curve under bending for columnar model ice. The legend also contains the stiffness ratio between the individual piecewise linear section and the "linear approach" (Gralher 2017)*

von Bock und Polach and Molyneux (2017) introduce the concept of plasticity ratio, *pr*, in order to signify the deviation of the actual strain modulus, *S*, from the elastic modulus, *E*. As Equation (1) indicates, *pr* expresses flexural plasticity;

$$pr = \left(1 - \frac{S}{E}\right) 100\% \qquad [1]$$

Values of *pr* were calculated as being 80% for data in von Bock und Polach et al. (2013) and 35% to 57% for the data in Enkvist and Mäkinen (1984). In Sazonov and Klementeva (2011) the elastic modulus derived from the force-displacement curve (e.g., *Figure 4*) led to a lower value than that obtained using the Hertz method (*Table 2*) and, consequently, a lower value of *pr*.

The initial slope pf the load-deformation curve reflects the reduced stiffness attributable to the early onset of plastic deformation or root deflection. Comparatively soft, compliant ice sheets are prone to be of low stiffness and have a low yield limit (von Bock und Polach et al. 2013). If the yield limit is reached directly during the load application, the *pr* value can be low because the initial stiffness, with a significant plastic contribution, would differ less from the global stiffness (Figure 5). In accordance with Equation 1, this behavior implies that, values of *pr* should not be based on time histories of flexural force.

Consequently, this paper introduces the term Non-Linearity of Model Ice (NLMI). NLMI is calculated similarly as *pr* (Equation 1). However, now the slope from start to end of the force-deflection curve represents the sheet's global stiffness, $S_g$, as estimated assuming the linear approach proposed by Gralher (2017). When first deflecting an ice-sheet downwards, the sheet's initial stiffness exhibits elastic linearity (e.g., at i = 1 in Figure 5), $S_{int}$, in Equation 2;

$$NLMI = \left(1 - \frac{S_g}{S_{int}}\right) 100\% \qquad [2]$$

The ensuing sections of this paper use NLMI as an indicator of plasticity. Provided model-ice stiffness (or deformation slope) changes, NLMI exceeds zero; thereby indicating structural changes in the model ice sheet. An example is presented in Figure 6 for NLMI = 91%, estimated using Equation 2 with $S_g = 1$ and $S_{int} = 11.4$.

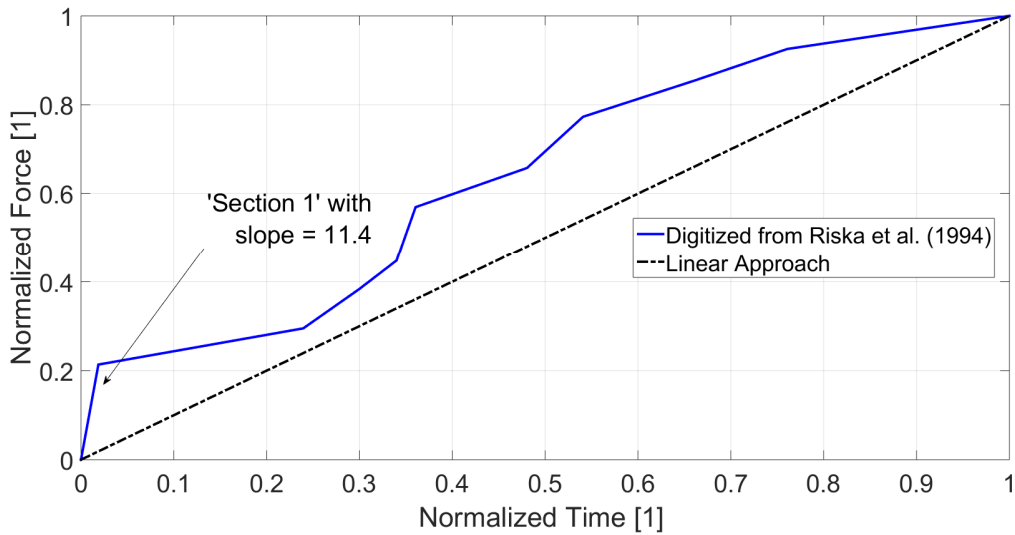

*Figure 6: Normalized force-time plot of a digitized cantilever beam test (Figure 17) from Riska et al.*
*(1994) and the Linear Approach proposed by Gralher (2017).*

## 3.2. Variation of strength behavior with sheet thickness

Model ice is usually treated as a homogeneous material (ITTC, 2014). However, this assumption
does not actually hold, because columnar model ice consists of two layers of different ice-crystal
structure (Figure 2a). The top layer has randomly oriented ice crystals, and is substantially the
stronger layer, as *Figure 7* illustrates from Grahler's (2017) estimates of the pressure distribution
of ice sheet in compression.

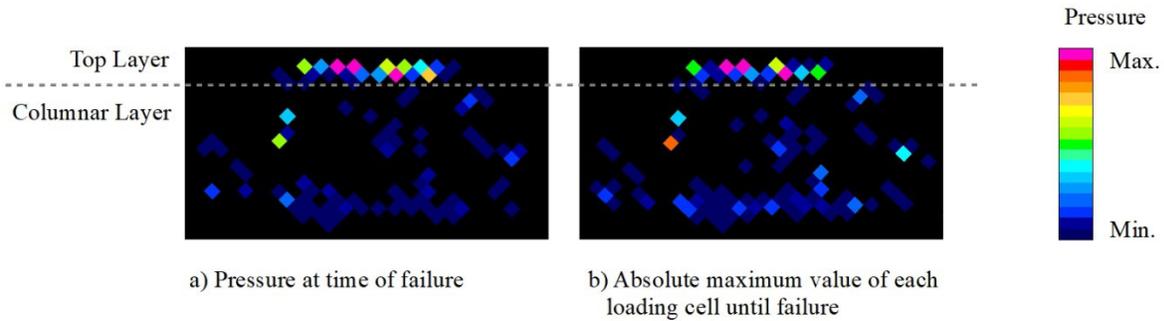

*Figure 7: Pressure distribution in columnar-grained model ice under uniaxial compression with columnar model ice doped with sodium-chloride (Gralher 2017).*

The mechanical properties of sea ice also vary from top to bottom of an ice sheet, in line with the temperature gradient through the ice (Assur 1967, Kerr and Palmer 1972, Kujala et al. 1990). However, the gradient is much less than that through a sheet of model ice (von Bock und Polach 2015, 2016).

The difference in properties of the two layers of the columnar ice prompted the development of fine grained (FG) model ice, a model ice intended to be of more uniform ice-crystal structure (Enkvist and Mäkinen 1984). However, FG model ice also forms a harder top layer (Figure 2b), who's thickness coincides with the freeboard of the ice (von Bock und Polach 2015). To reproduce the downward flexure behavior of ice (von Bock und Polach 2015), it was found that $E$ had a significant gradient (Figure 8) through the ice thickness. Valkonen et al. (2007), at Aalto University (using FGX model ice), found the same trend.

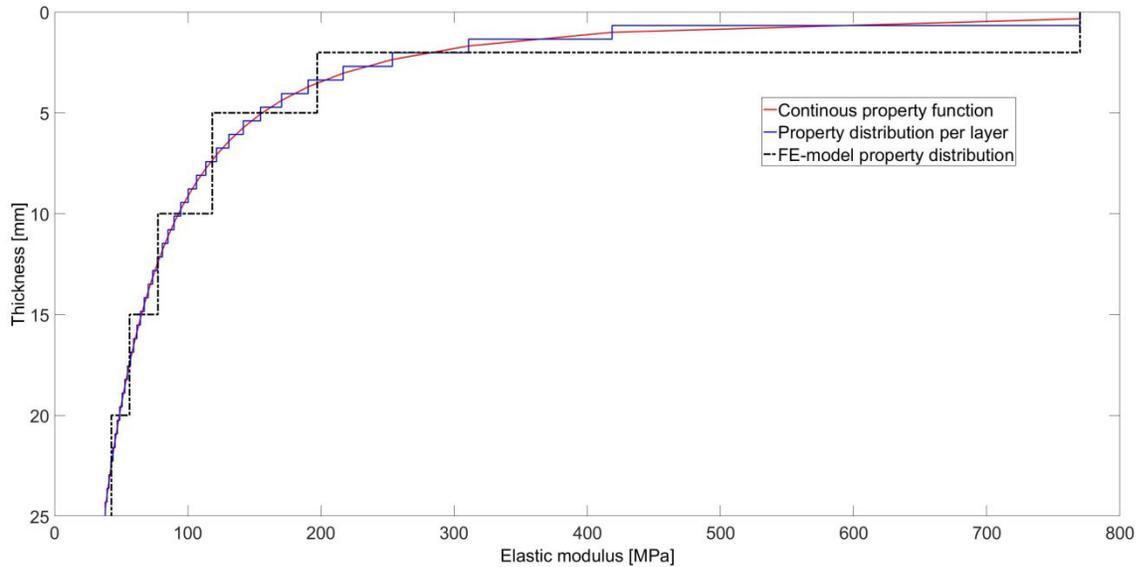

*Figure 8: Variation of elastic modulus, E, through a FG model-ice sheet (von Bock und Polach, 2015), determined by the means of experiments (von Bock und Polach et al. 2013) and a numerical model (von Bock und Polach 2015)*

## 4. Ice tank data

Data on the downward-flexure behavior of model ice sheets used at the leading ice tanks exhibit distinctly non-linear, flexural behavior. The present study, moreover, finds that all aqueous forms of model ice exhibit this behavior. Taken from information given by the leading ice tanks (Table 3) and published articles, the data cover almost the entire range of model ice types. Only the model ice sheets used by NRC and KRISO are not included among the data, as their standard measurement procedures do not produce the force-time history plots needed for the evaluation.

*Table 3: Origin and properties of data used in the analysis. If the origin of data contains the name of the ice tank, then those information are not publically available and are provided by the ice tank directly for the here presented research.*

| Ice Tank | Source | Grain Structure | Dopant (% of weight) | Thickness (mm) | Flexural strength (kPa) |
|---|---|---|---|---|---|
| AORC (now Aalto Ice Tank) | Jalonen et al. (1990), Riska et al. (1994), Li and Riska (1996) | FG | Ethanol (0.5 %; 0.1 %) | 30 - 49 | 43 - 50 |
| Aalto Ice Tank | von Bock und Polach (2015) | FG | Ethanol (0.3 %) | 25 - 27 | 50 - 77 |
| The Hamburg Ship Model Basin (HSVA) | Evers and Jochmann (1993), Gralher (2017), HSVA (2018) | Columnar | NaCl (0.9%, 0.7 %) | 19 - 52 | 15 - 79 |
| Krylov State Research Centre (KSRC, old ice tank) | Sazonov and Klementeva (2011) | FG | NaCl (1.5 %) | 12 - 59 | No data available (NDA) |
| | | Columnar | NaCl (1.5 %) | 10 - 40 | |
| Krylov State Research Centre (KSRC, new ice tank) | KSRC (2018) | FG | NaCl (1.5 %) | 26 - 87 | 14 - 39 |
| Tianjin University | Huang et al. (2018) | Columnar | Urea (1.5 %) | 40 | 32 |
| National Marine Research Institute (NMRI) Japan | NMRI (2018) | Columnar | Propylene Glycol (0.8 %) | 15 - 17 | 41 – 76 |
| WARC (now Aker Arctic) | Soininen and Nortala-Hoikkanen (1990); | FG | NaCl | No data available, only force-time plot | |
| Aker Arctic (AARC) | AARC 2018 | FG | NaCl (1.1 %) | 27 - 73 | 31 - 49 |
| University of Iowa | Cook (1983) | Columnar | Urea | Measurements on temperature and urea-brine | |

All the data sets have force-time or displacement plots, and include information such as ice-sheet thickness, flexural strength, dopant concentration and others. However, the level of completeness of the datasets varied, thereby excluding some data from certain analyses conducted for this study. In total, 66 data sets were available, of which 40 sets were for FG model ice and 26 were for columnar model ice. *Figure 9* plots the experiments for which ice thickness and flexural strength are known. There are 13 data sets for which thickness or flexural-strength data are missing. The force versus time, or force versus displacement, plots available only as figures were digitized for the analysis.

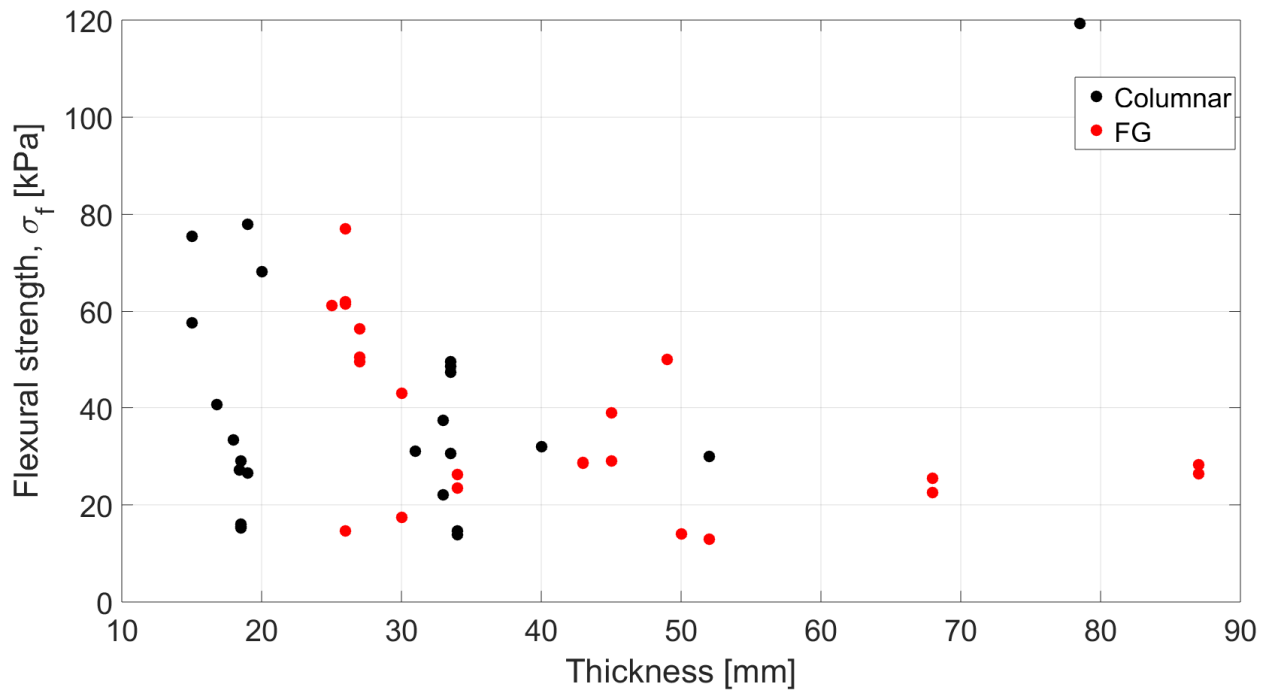

*Figure 9: Columnar and fine grained (FG) force-time data of flexural strength measurements for which thickness and flexural strength are available.*

## 5. Data Analysis

The study's analysis related values of NLMI to characteristic properties of the model ice sheets, in order to assess the importance of several factors influencing the non-linear behavior. For this purpose, the Pearson correlation coefficient (Pearson R test), $r$, was used. As defined in Equation 4, $r$ is a common method to quantify univariate, linear correlation between two sets of measurements;

$$r = \frac{\sum(x_i - \bar{x})(y_i - \bar{y})}{\sqrt{\sum(x_i - \bar{x})^2 \cdot \sum(y_i - \bar{y})^2}} \qquad [4]$$

Here, $x_i$ and $y_i$ are the measurements or observations, and $\overline{(\cdot)}$ is the mean of the respective set of measurements. The value of $r$ varies between $-1$ and $1$, where the highest absolute value indicates a perfect correlation. For an $r$ value close to one, the majority of data points fit along a diagonal. Values of $r$ close to zero indicate no linear correlation, but non-linear relationships between two sets might still exist (Gibbons, 1986). A guideline for the interpretation of $r$ is given in Table 4; correlations with $r$ values lower than 0.29 were discarded.

*Table 4: Interpretation of* analysis *results of the Pearson correlation (ranges of r suggested by the Political Science Department at Quinnipiac University)[2]*

| Correlation coefficient, $r$ | Significance |
|---|---|
| +/- 0.70 to +/- 1 | Very strong relationship |
| +/- 0.40 to +/- 0.69 | Strong relationship |
| +/- 0.30 to +/- 0.39 | Moderate relationship |
| +/- 0.20 to +/- 0.29 | weak relationship |
| +/- 0.01 to +/- 0.19 | No or negligible relationship |
| 0 | No relationship (zero correlation) |

---

[2] http://www.statisticshowto.com/probability-and-statistics/correlation-coefficient-formula/, 13.09.2018

## 5.1.  Data description and data sets

Table 5: *Description of parameters* summarizes the data sets and their origins. The numerical

values of each parameter are found in Appendix A.2.

*Table 5: Description of parameters*

| Name in Pearson correlation analysis | Units | Explanation |
|---|---|---|
| plasticEnergy | [1] | Normalized energy expended in deformation process ( |
| maxDeflection | [mm] | Maximum deflection at the loaded beam tip |
| flexuralStrength | [kPa] | Measured flexural strength |
| deformationVelocity | [mm/s] | Deformation velocity at the loading point |
| loadingTime | [s] | Time from load application to beam failure |
| maxForce | [N] | Maximum load at point of failure |
| thickness | [mm] | Thickness of the ice |
| weightChemical | [g/mol] | Specific weight of the chemical / dopant used to adjust ice properties |
| dopantConcentration | [%] | Concentration of the dopant / added chemical related to the weight water (in all cases this reference is identical) |
| relativeWeight | [g/mol] | Specific weight of the chemical multiplied by the dopant concentration (in % of weight) |
| relativeVol | [cm³/mol] | Relative Weight divided by chemical density |
| bulkCoeff | [] | The bulk coefficient, ke, refers to table xx in Timco(1981) and reflects the amount dopant trapped within the ice relative to the amount in the water |
| KeWeight | [g/mol] | This is the relative weight multiplied with the bulk coefficient which is eligibly the actual amount trapped within the ice. |

The dopant variables listed in Table 6 were directly measured, except the *relative weight*, the

*relative Vol*ume and the *keWeight*. Dopant concentration is given as a percentage of the weight

of fresh water. The relative weight is the product of the specific weight and the dopant

concentration. Dividing this quantity by the density gives the relative volume. Timco (1981)

conducted experiments with ice frozen from impure melts and determined values of bulk coefficient, *ke*, reflecting the amount of dopant trapped within the ice relative to the amount in the water solution.

*Table 6: Dopant properties of analyzed sheets of model ice*

| Name | Chemical formula | Specific weight [g/mol] | Density [g/cm³] | Melting point [C°] | Effective bulk partition coefficient, *ke* (Timco 1981) |
|---|---|---|---|---|---|
| Salt (sodium chloride) | NaCl | 58.44 | 2.16 | 801 | 0.3 |
| Ethanol | CH3CH2OH => C2H6O | 46.069 | 0.789 | -114.1 | 0.87 |
| Urea/Carbarmide | CH4N2O => NH2CONH2 | 60.056 | 1.32 | 133 | 0.39 |
| Propylene glycol | C3H8O2 | 76.02 | 1.04 | -59 | unknown |

## 5.2.    Analysis of the causes of non-linear, flexural behavior

An analysis of causes of non-linear flexural behavior of model ice sheets was performed for the data, and then separately for the two types of model ice sheet: columnar ice; and, FG model ice. The purpose of the analysis was to assess the extents to which model ice sheets were influenced by their different crystal structures.

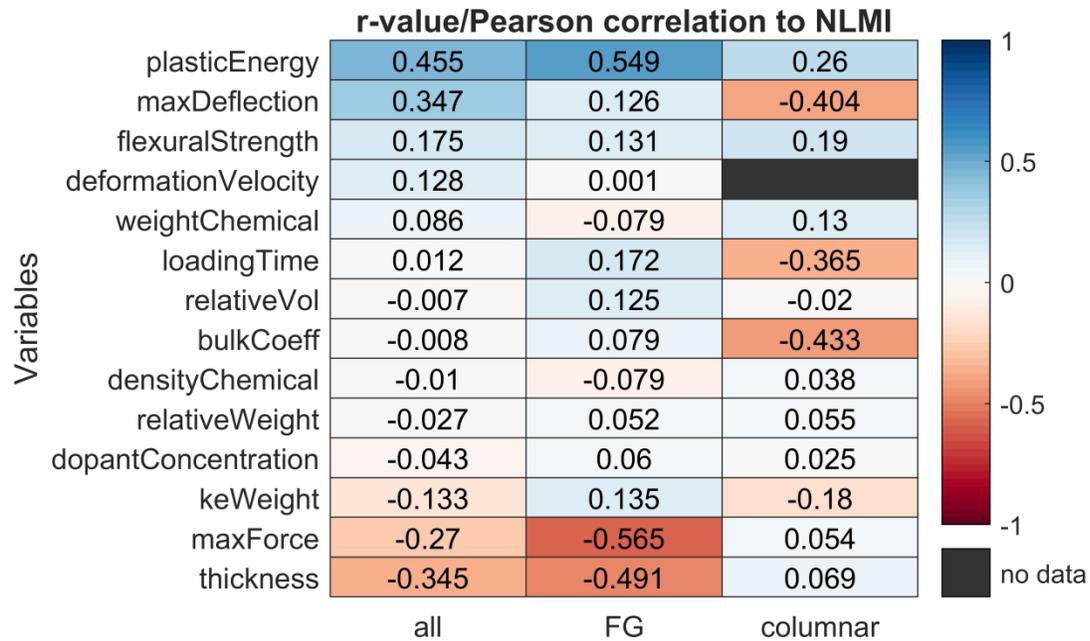

*Figure 10: Pearson correlation (r value) for NLMI evaluations for all the data and a separate analysis of only columnar and only fine-grained model ice.*

Table 7 summarizes the three most significant variables for the different analyses that have values larger than 0.3 or smaller than -0.3. Also, Table 7 gives the number, $n$, of samples used to estimate the $r$ value.

*Table 7: Summary of the most influencing features to the NLMI from the Pearson correlation.*

*The corresponding value of r as well as the number of samples (n) is found in brackets.*

| Analysis | No. of samples | Significant Parameters (Correlation at least moderate) | |
|---|---|---|---|
| | | Positive correlation | Negative correlation |
| *Full data set* | 66 | 1. *Plastic energy (PCC = 0.46, n = 66)*<br>2. *Maximum tip deflection(PCC = 0.35, n = 24)* | 1. *Ice thickness (r = -0.35, n = 64)* |
| *Only columnar model ice* | 26 | *(no entries)* | 1. *Bulk coefficient, ke (r = -0.43, n = 23)*<br>2. *Maximum tip deflection (r= -0.4, n = 3)*<br>3. *Loading time (r = -0.37, n = 25)* |
| *Only FG model ice* | 40 | 1. *Plastic energy (PCC = 0.55, n = 40)* | 1. *Maximum tip force (r = -0.56, n = 21)*<br>2. *Ice thickness (r = -0.49, n = 38)* |

# 6. Presentation of Results

This section presents results giving insights into the physical causes linking the variables in Table 7 with the resulting value of NLMI.

## 6.1.    Expended elastic-plastic energy

Expended elastic-plastic energy is defined by the area under the force-time curve (Figure 6). Analysis of all the data shows a strong positive correlation of the NLMI with the normalized elastic-plastic deformation energy.  In Figure 6 , the presumed yield strength is around the value of 0.2. If the yield strength were moved to a higher force level, the expended elastic-plastic deformation energy also increases. This process occurs when the deformation progression between the initial yield point and failure is strongly hyperbolic; i.e., the curve has a strongly developed gibbous hump.  A strongly hyperbolic progression can occur when an ice sheet has several layers differing in crystal structure and varying in strength from top to bottom (Figure 8 with six layers). Owing to the non-uniform stress distribution through thickness in a deflected cantilever beam, the different layers yield at different load instances. With the yielding of each layer, the global stiffness reduces and, thereby the slope of the time-force continuously reduces (see von Bock und Polach, 2015 and 2016 for reference).

The full set of data, and data for only the FG model ice, show strong positive correlation with elastic-plastic energy, whereas the data for only the columnar model ice show weak positive correlation. FG model ice has a homogeneous texture through which a continuous functional grading of properties can occur; some properties may still vary through each crystal layer from top to bottom. The more layers of ice undergo the transition from elastic to plastic behavior, the more hyperbolic the force-time curve becomes and, consequently, the deformation energy increases. The two distinctly different crystal layers of columnar ice cause a more discrete distribution of properties through thickness and consequently a less pronounced hump in the force-displacement curve.

## 6.2.     Loading time

For only columnar ice, unlike the result for the full set of data, loading time (and therefore strain rate) had a strongly negative correlation with NLMI. This finding indicates that NLMI decreases with increasing loading time. For sea ice or prototype ice, the impact of loading time (strain rate) is well established (Sanderson 1988). Below a critical strain rate, ice cannot be considered to respond elastically all the way to complete brittle failure, as idealized to occur when ship hulls break ice sheets. However, the strain rate sensitivity of ice in tension is significantly less pronounced than in compression (Timco and Weeks, 2010). As critical strain rates are not yet defined for model ice, the correlation in Table 7 seems contrary to sea ice behavior; further explanation is needed.

## 6.3.    Maximum tip deflection

The moderately positive correlation of maximum tip deflection with NLMI for the full data set is an expected correlation. The value of NLMI reflects the degree of plasticity of the model ice and consequently an exaggerated large tip deflection due to occurring plasticity (Schwarz, 1977) has to correlate positively with the NLMI. For columnar ice, an increase in tip deflection results in a smaller value of NLMI. This finding seems counter-intuitive and opposite to what Schwarz (1977) and other express; i.e., excessively large (not-scalable) deformations caused by plasticity should be avoided, because they increase sheet resistance.

But, the data sets pertain to ice sheets of various thickness and strength combinations and cantilever beam tests in ice tanks are conducted in accordance to ITTC guidelines (ITTC, 2014). The guidelines recommend that tested cantilever beams fail within 1s to 2s. This criterion evidently was met for all analyzed beams.  However, the guidelines also require cantilever beams whose length is 5-7 times sheet thickness. Consequently, the thicker ice, the longer the beam. Accordingly, the longer a beam, the larger is the deflection of its tip (assuming the same flexural strength). Additionally, the cantilever beam tests are maybe affected by displacements at the beam's root into the sheet.  This effect causes an initial bending angle at the beam's root that increases the displacement with increasing beam length.  Therefore, an increase in maximum tip displacement can be associated with both increasing plasticity of sheet and increasing ice thickness.  The latter variable correlates strongly negatively with the NLMI, as the following section addresses.

## 6.4.    Ice thickness

Ice-sheet thickness correlates negatively with NLMI values, as NLMI increases for thinner ice sheets. This result is especially significant for FG ice, as Figure 11 shows.  Plotted values of NLMI are averaged over 20 mm thickness bins, with error bars reflecting the standard deviation of the values. Columnar model ice has only two data points beyond 40 mm thickness and has therefore an *r* value close to zero.

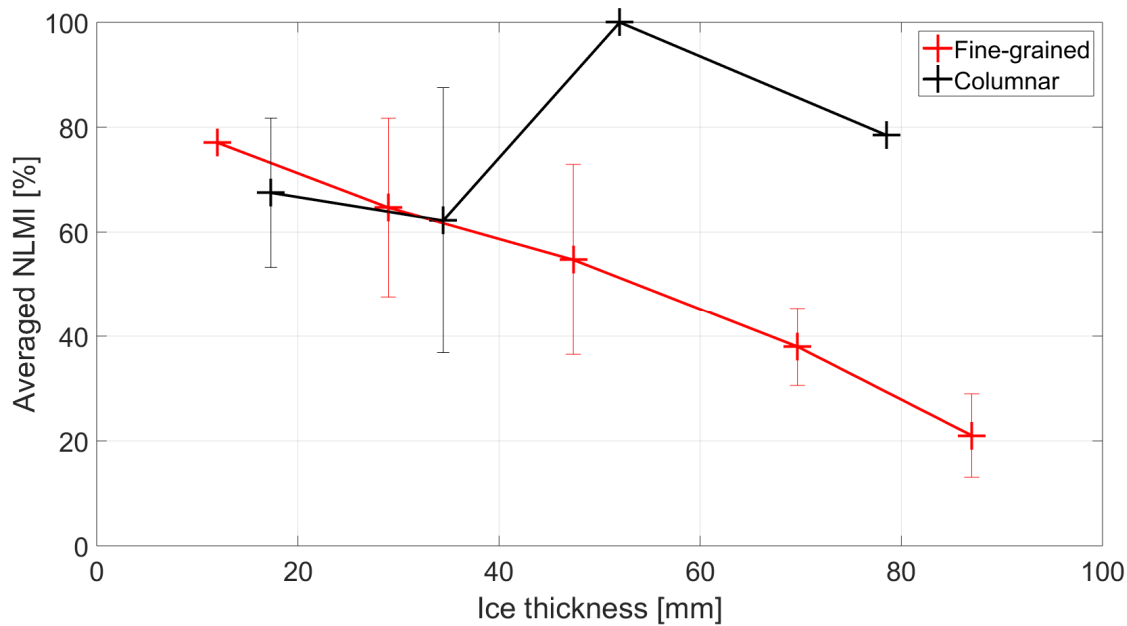

*Figure 11: Average NLMI values for fine grained and columnar ice.  Each marked point reflects the averaged thickness and NLMI values over the ranges marked by the grind lines of the horizontal axis. The error bars reflect the standard deviation i.e. the scatter at the particular points.*

Further, Figure 12 shows that sheet thickness affects NLMI, as a low value NLMI occurs for the thicker ice and a higher value of NLMI occurs for the thinner ice. The flexural strength for the sheets was almost identical, but the crystal structures of the sheets differed.

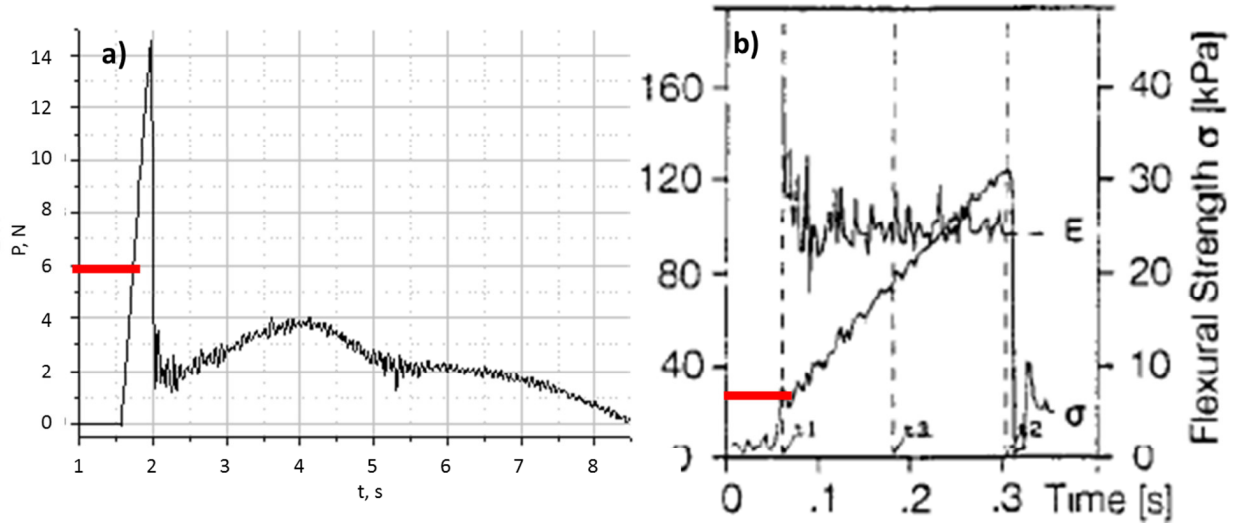

*Figure 12: a) Force-time curve of a cantilever beam test of a granular NaCl doped ice, in 87 mm thick ice with a flexural strength, $\sigma_f = 28,2\ kPa$ and a low non-linearity, NLMI, of 15%, (measured at KSRC, 2018). b) Force-time curve of a cantilever beam test of columnar NaCl doped ice, in 31 mm thick ice with a flexural strength, $\sigma_f = 31\ kPa$ and a high non-linearity, NLMI, of 95% (from Evers and Jochmann 1993).*

The red bars in in Figure 12 indicate the anticipated yield point; i.e., where the initial stiffness of a sheet changes. In Figure 12a, the point of stiffness change can be considered arbitrary, because the change is not pronounced and the sheet fails brittle in a brittle manner without significant prior deformation yielding.

Sheet strength and stiffness distributions are non-linear through sheet thickness, especially for relatively thick ice. Further, thicker ice sheets have layers of different crystal structure (Figure 8). However, calculation of the bending strength (following ITTC, 2014) assumes a homogeneous distribution of properties from top to bottom of the sheet. In practice, the small strength contribution of the bottom layer has to be compensated by the strength contribution of the top layer. Therefore, the strength and flexural stiffness of the relatively thin top layer need to be substantial. von Bock und Polach (for the sheet in Figure 8) show that the strength of the top layer is about 60% of the sheet's total flexural strength, which can be close to the strength of prototype ice for thick ice and therewith also other properties such as elastic modulus or elastic limit.

## 6.5.    The dopant

Dopant solutes are used to adjust the strength of model ice. Timco (1986) and Cole (2001) state that high concentrations of dopant solute can produce excessively large plastic deformations. Therefore, NLMI values should correlate positively with certain characteristics of dopants; e.g., increased concentration of dopant increases the non-linear flexural behavior (and NMLI value) of ice sheets. Figure 13 shows this trend for ethanol doped ice sheets formed in the ice tank of Aalto University. The trend, though, does not account for the influence of sheet thickness or flexural strength, and thereby the variation of brine volume within the sheet.

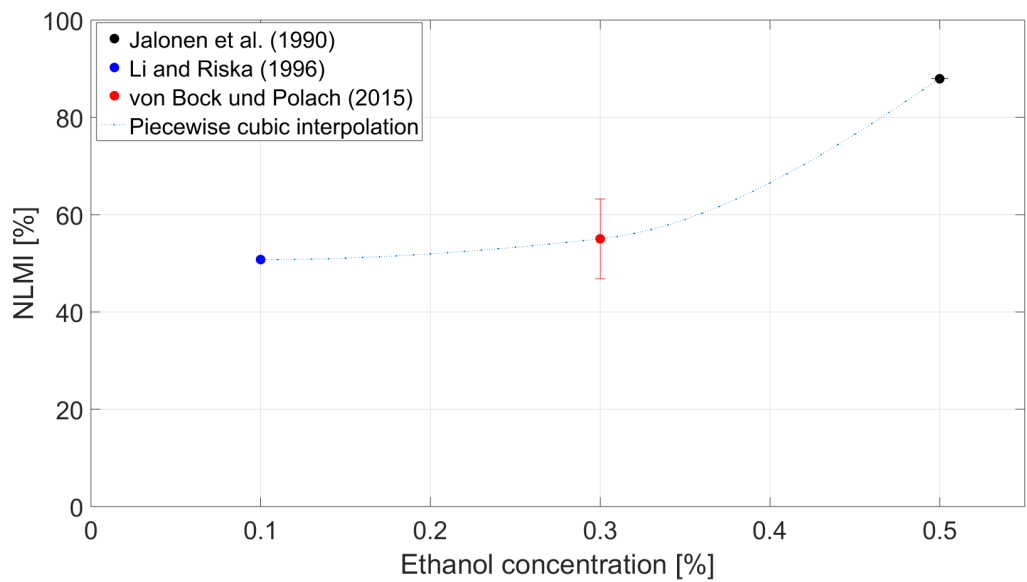

*Figure 13: NLMI in dependence on the concentration of the dopant ethanol in tank water.*

A useful variable for characterizing solute content is bulk coefficient, *ke*, the ratio between the amount of solute trapped within the ice and the amount in the solution. According to Timco (1981), values of *ke* vary significantly depending on the chemical used (see also Table 6). Timco (1981), though, determined values of bulk coefficient using ice samples in a different way that used currently for forming sheets of model ice. The *ke* value for NaCl is 0.3 (Timco, 1981 and Table 6), but in direct measurements at HSVA *ke* is actually between 0.35 and 0.42 (Table 8).

*Table 8: Measured values of ke for the columnar ice of HSVA at a NaCl for different combinations of ice thickness and flexural strength. Values for thickness and flexural strength are averaged.*

| Thickness [mm] | Flexural strength [kPa] | Measured *ke* | NaCl dopant solute in tank water (% of weight) |
|---|---|---|---|
| 19 | 29 | 0.37 | 0.7 |
| 20 | 73 | 0.42 | 0.73 |
| 34 | 48 | 0.35 | 0.7 |
| 79 | 119 | 0.33 | 0.7 |

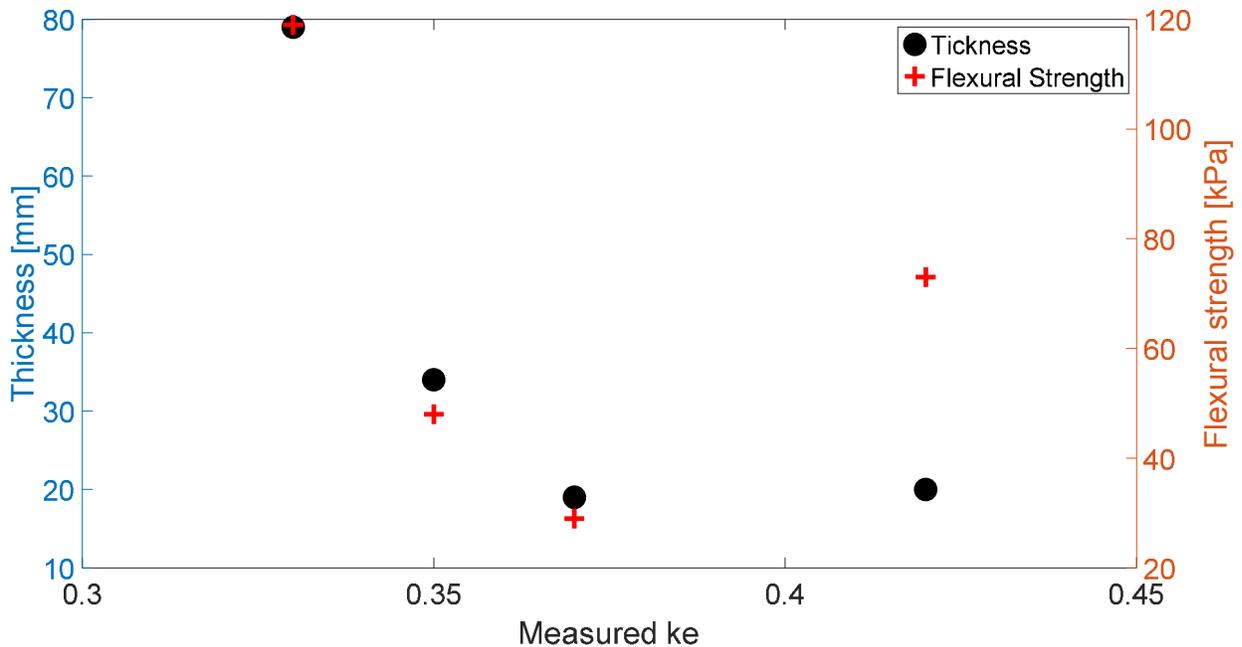

*Figure 14: Dependency of the measured bulk partition coefficient (ke) on the ice thickness and the flexural strength for columnar model ice grown from saline doped water (here data from HSVA).*

Figure 14 gives data indicating how values of *ke* vary with sheet thickness and flexural strength. This figure shows that *ke* values determined by Timco (1981) are not accurate for columnar model ice; presently, comparable measurements are not available for FG model ice. The growth process of the ice from Timco (1981) and columnar model ice is similar, apart from the seeding process. The production of FG ice differs and during growth no dopant is repelled, but all contents of the sprayed tank water are stored in each sprayed layer of the ice. Therefore, values of *ke* should be higher for FG ice.

The measurements in Figure 14 indicate that the effective value of *ke* does not accurately indicate the strength state of model ice. This observation agrees with the measurements in Figure 15, which shows that the unlocking of brine (dopant solute) content is a dynamic and possibly non-linear process involving the growth of brine volumes within an ice sheet. Cook (1983) gives data supporting this point. He measured solute content in a 30 mm-thick sheet of urea ice during tempering (warming) of the sheet. Figure 15 shows the measured solute content in percentage by weight commensurate with the warming cycle in Figure 16 (Cook 1983). While the increase of temperature within the ice is approximately linear the increase or unlocking of solute is non-linear and accelerated. Tempering the model-ice sheet, therefore alters the temperature distribution through the sheet, and increases the volume of voids containing concentrated solute between the ice crystals. The increase of solute correlates well with the reduction of strength.

Tempering the model-ice sheet alters the temperature distribution through the sheet, and increases the volumes of voids with concentrated solutes between the ice crystals and the other way around. Increased warming and therewith an expansion of brine is expected to increase the plasticity in analogy to sea ice (Cole 2001).

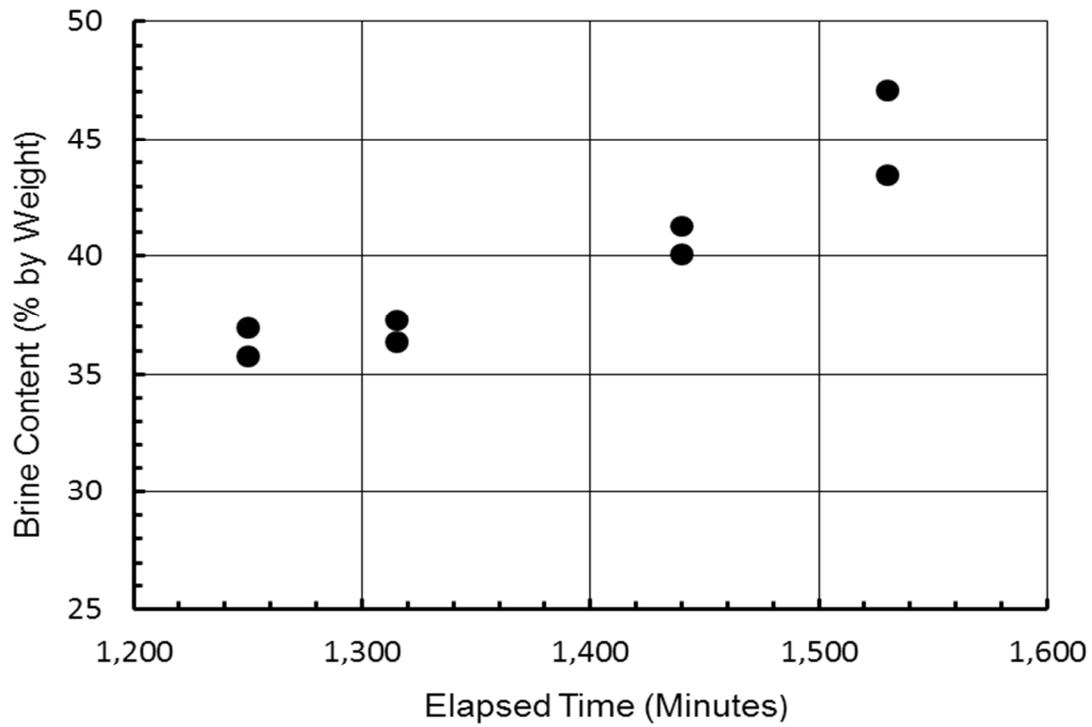

*Figure 15: The time variation of urea brine content in ice beams obtained from the ice sheet during the warming period, which began at zero minutes. The two measurement points refer to two examined ice beams cut from the ice sheet (Cook 1983).*

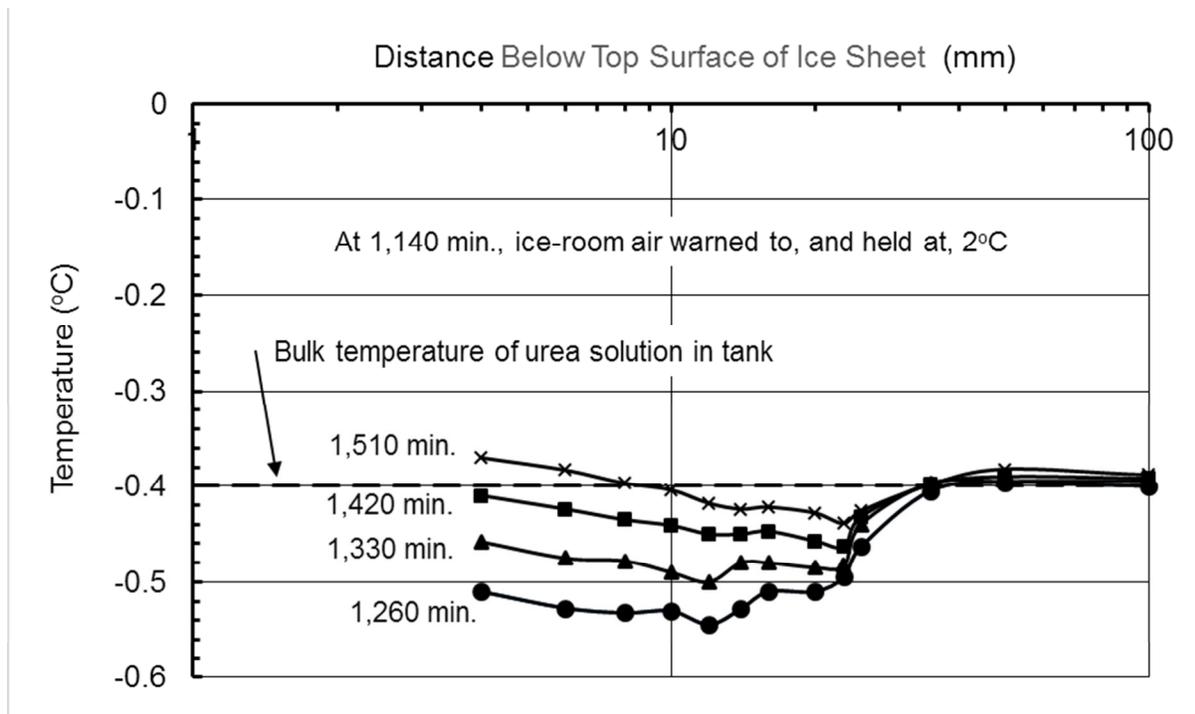

*Figure 16: The warming of the same ice sheet as in Figure 15 was with 2°C room temperature from 1,140-1,520 minutes for the 30 mm thick ice sheet (Cook, 1983).*

## 7. Discussion of results

Between 1975 and 1990, much research was done to develop model ice to better meet similitude constraints. The non-linear flexural behavior of model ice stimulated much of the research (e.g. Schwarz, 1977, Enkvist and Mäkinen, 1984, Evers and Jochmann, 1993).  Excessive, flexural deformation of model ice led to concerns for scale effects expressed as exaggerated vertical motions of ship hulls moving through ice sheets. The present analysis (building on work by von Bock und Polach, 2015) shows that all current forms of aqueous model ice exhibit non-linear, flexural behavior with plasticity substantially affecting flexural behavior.  The NLMI (Equation 2) parameter was developed to improve characterization of the elastic or elastic-plastic response of ice sheets in downward flexure.

The linear correlation analysis presented above confirms that the strength properties of the top layer of model ice play a significant role in the deformation process of FG ice (see also von Bock und Polach, 2015), which was generally considered more homogeneous in crystal structure than columnar ice (Enkvist and Mäkinen, 1984).

An abiding fundamental assumption is till that sheet plasticity is strongly affected by solute voids, as it is with sea ice (Cole, 2001). However, no linear correlation has been found for the model ice data analyzed here. The work of Cook (1983, Figure 15) shows a non-linear increase of unlocked brine weight with increasing temperature. Moreover, Figure 16 also shows a non-linear relationship for $ke$ relative to sheet thickness and flexural strength for the HSVA data. The solute content within a sheet is subject to unsteady, non-linear processes, but the parameters used in the present correlation analysis are constants based on Timco (1981). This difference indicates a significant gap of knowledge. Consequently, the negative correlation of $ke$ values with NLMI values for columnar model ice (Table 7) should be viewed with caution, as this result may be coincidental and currently cannot be attributed to physical processes. A weakness of the Pearson correlation approach is that it may inadequately reveal the net effect of multiple processes occurring together; dopant solute behavior, ice thickness effects and overall non-linear behavior.

However, the analysis shows that the top layer of model ice sheets play a major role in the deformation of model ice and that, for thick ice, values of top-layer strength and stiffness can be comparable to values for prototype ice. The top layer is directly exposed to cold air and, therefore, has a significantly lesser amount of solute liquid than the warmer bottom layers, in which temperature approaches the melting point of water. If the top layer dominates the deformation behavior of model ice, the following question arises: How significant for sheet deformation behavior are the bottom layers and the contained brine liquid? It is possible that dopant solute volumes have a relatively low physical significance in the overall flexural behavior of model ice sheets; the NLMI values infer this finding. Additionally, the relatively strong top layer limits modelling thin ice, because the strength properties of model ice are too high compared to strengths prescribed by model length scale.

An interpretation of the analysis is that the non-linear behavior of model ice appears not to depend strongly on the crystal structure of the ice and or the dopant. The model ice sheets formed at all the ice tanks showed a non-linear flexural behavior. This interpretation infers that the ratio $E/\sigma_f$ plays a relatively insignificant role if $E$ is determined mainly using the Hertz method (ITTC, 2014). Preferably the strain modulus should be determined using the cantilever beam test method (ITTC guideline 7.5-02-04-02, Section 3.2), so as to give an effective value of sheet stiffness that a ship experiences when deflecting and flexurally breaking an ice sheet.

The data analyzed in this paper were collected from the world's leading ice tanks and include the two main crystal-structure forms of model ice sheet, various dopants and dopant concentrations, and a range of ice-sheet thicknesses and flexural strengths. Nonetheless, despite the large amount of data analyzed, some combinations of ice sheet thickness and strength were found to be fairly scarce: flexural strength values beyond 60 kPa; sheet thicknesses greater than 30 mm; and, flexural strength values for sheet thicknesses greater 60 mm (Figure 9).Furthermore, identical strength-thickness combinations of different model ices would be needed for distinct analysis of the impact of the dopant and the crystal structure on the non-linear, flexural behavior.

## 8. Conclusions

This paper analyzes the non-linear, flexural behavior of the two main forms of aqueous model ice sheet (columnar and fine grain) used to simulate the downward flexure and failure of ice sheets during ship hull passage through an ice sheet. Data on this behavior were obtained from the world's leading ice tanks (Table 3).

The analysis, which used the Pearson correlation coefficient test to evaluate the influence of variables, showed that strong similarities exist in the flexural behavior of all model ice types, though the types involve different ice-crystal structures and dopants. All forms of model ice sheet were found to exhibit non-linear, flexural behavior. Such behavior is evident as significant plasticity of model ice in the latter stages of downward flexure resulting in ice breaking. The degree of non-linearity, herein expressed using the parameter NLMI, depended strongly on sheet thickness.

The analysis was inconclusive as to the extents to which dopant presence (type and concentration) influenced the non-linear, flexural behavior of model ice sheets. Dopant presence affects this behavior because dopant presence affects crystal size, and the extent of dopant solute affects sheet strength during sheet warming or tempering. Ice-tank data are presented showing this effect. However, the applied methods for estimating ice-sheet strength behavior (notably the Hertz method mentioned below) can affect values of reported data. Further, sheet thickness introduces non-linearity in flexural behavior, thereby affecting reported data.

The non-linear behavior of model ice complicates satisfying the requirements of Cauchy number similitude, expressing the ratio of inertia to elastic-deformation forces. The analysis shows that values of the ratio $E/\sigma_f$ should not be calculated using the sheet's elastic modulus as determined via the Hertz method. Instead, the use is recommended of the average strain modulus determined by means of cantilever beam deflection prior to conduct of modeling (ITTC guideline 7.5-02-04-

02, Section 3.2). This measurement reflects the actual flexural deformation behavior of a sheet of model ice.

Despite the variation of strength properties through sheet thickness, and the overall non-linear behavior in downward flexure, existing forms model ice sheets work suitably well for physically modeling most scenarios of flexural failure of ice sheets, such as during hull passage. However, the difficulty to comply with similitude targets indicates that it may not be feasible to produce a model ice material that reproduces all aspects of sea ice. behavior.

A practical approach in future ice-tank modeling would be to use different model ice materials or types for different modeling purposes. This approach, termed "case-based scaling" (von Bock und Polach and Molyneux, 2017) is already being implemented at some ice tanks. For example, HSVA is in the process of developing a new model ice for the compressive failure of ice against a vertical structure (Ziemer, 2018).

The analysis points to the great value of the experience and skill set of the modelers operating ice tanks and personnel operating actual icebreaker ships. The analysis also points to the ongoing need of further understanding the strength and deformation behavior of actual ice. For example, as tempering (warming) or weakening of an ice sheet is an unsteady and non-linear process, the constants as presented by Timco (1981) inadequately describe the solute content and influence, and suggest the need for more research regarding the processes involved in sheet tempering.

# Acknowledgements

The presented research was supported with measurement data provided by Aker Arctic (T. Leiviskä), HSVA (N. Reimer and G. Ziemer), Krylov State Research Centre (A. Dobrodeev) and the National Marine Research Institute of Japan (T. Matsuzawa). Their support is gratefully acknowledged.

# Appendix

## A.1.    Data extraction

Figure 17 illustrates a measurement from Riska et al. (1994), with marker points for digitalization by converting pixels into length scales.

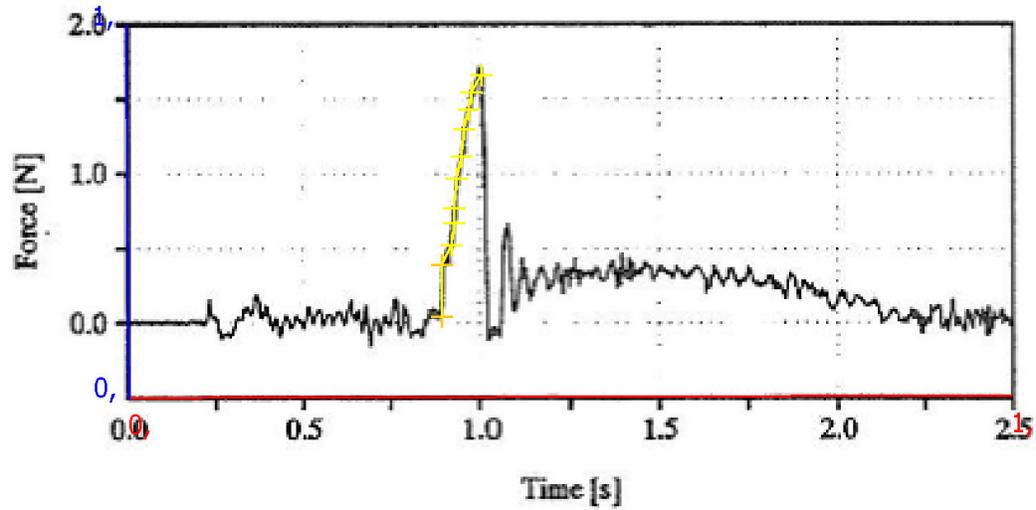

*Figure 17: Force-time plot of a cantilever beam test with FG-X ice taken from Riska et al. (1994) after plot digitalization (with yellow markers).The ice thickness is 30 mm and the flexural strength 43 kPa.*

## A.2.    Data Sheet

An excel table is attached that contains the compiled data used in the analysis.

## A.3.    Plots for Pearson correlation

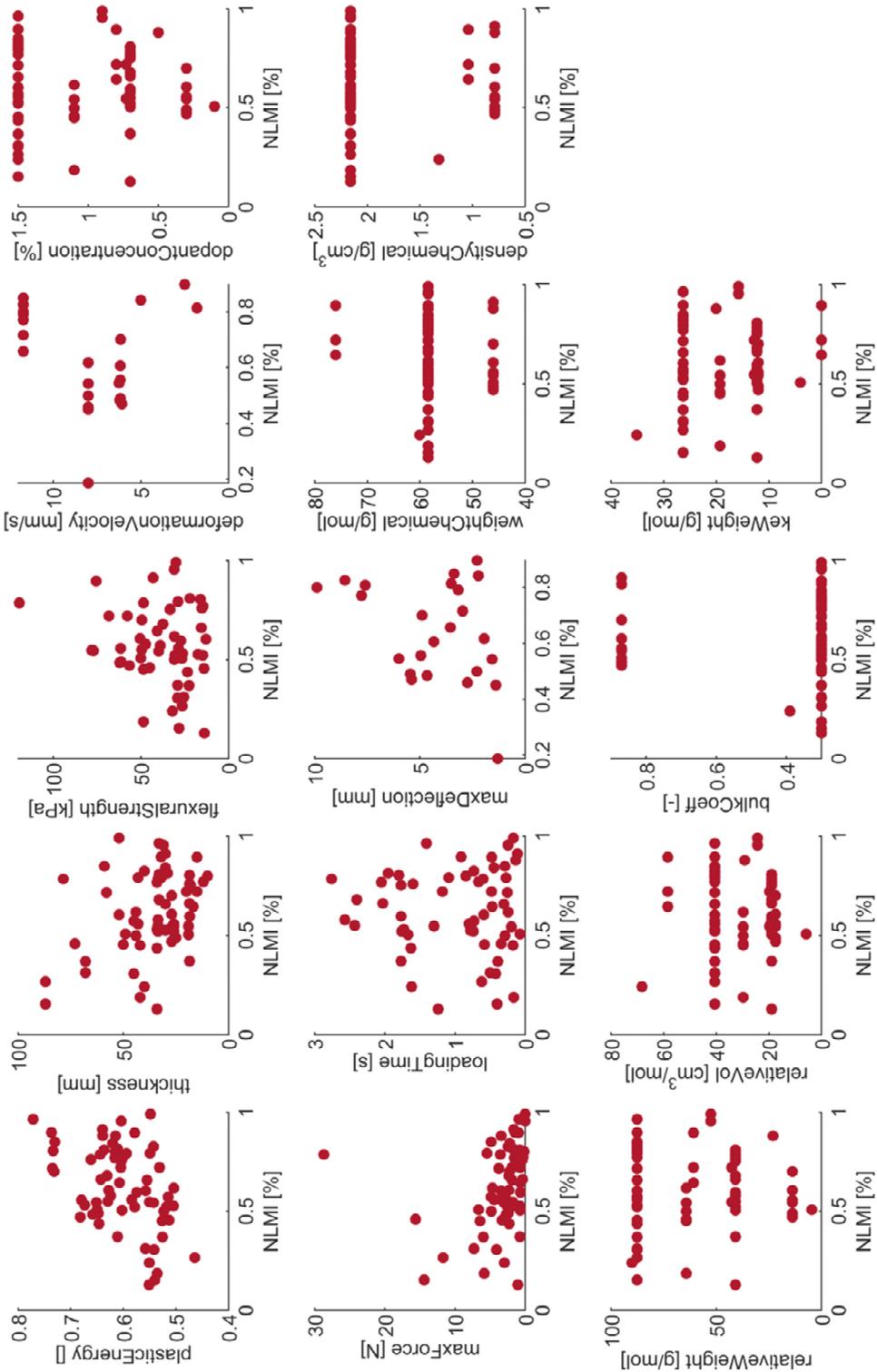

*Figure 18: Pearson correlation to NLMI for model ice (full data set)*

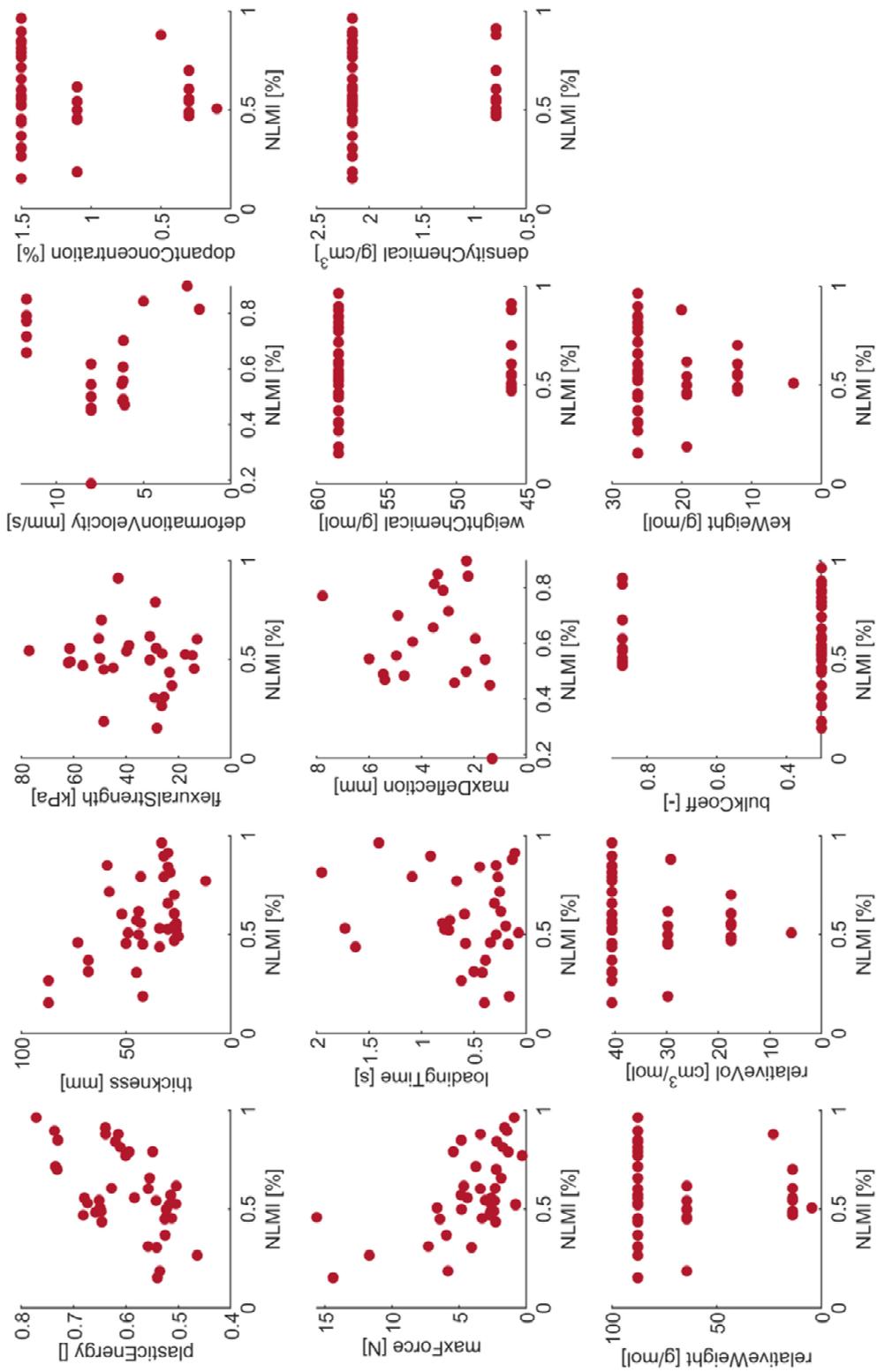

*Figure 19: Pearson correlation to NLMI for model ice with fine-grained structure*

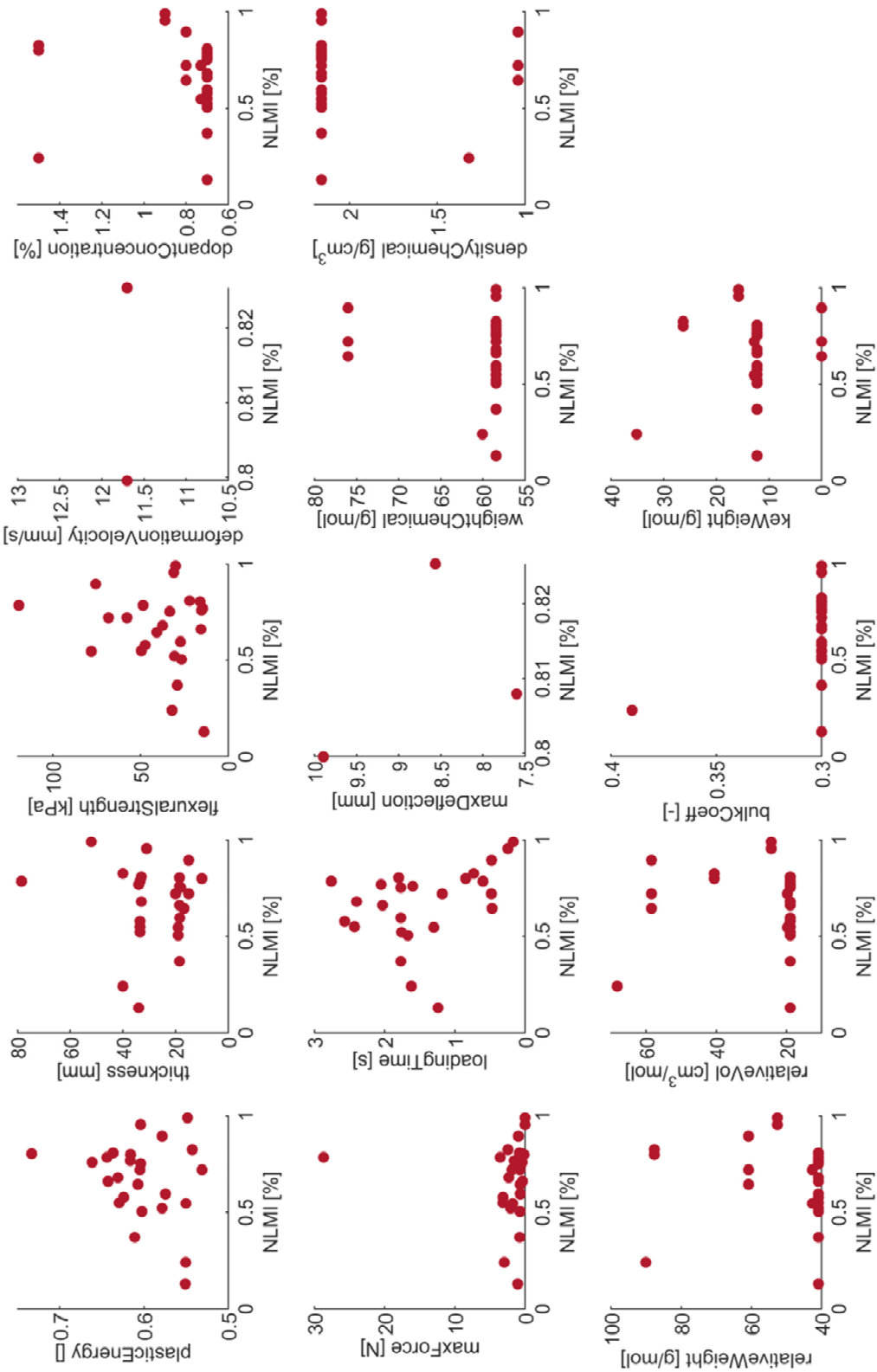

*Figure 20: Pearson correlation to NLMI for model ice with columnar structure*